\documentclass[sigconf]{acmart}
\usepackage{graphicx} 
\usepackage{tabularx}
\usepackage{subcaption} 
\usepackage{array,multirow}
\usepackage[noend]{algorithm}
\usepackage[noend]{algpseudocode}
\usepackage{xcolor}

\usepackage{amsmath,amssymb,amsfonts}

\AtBeginDocument{%
  \providecommand\BibTeX{{%
    \normalfont B\kern-0.5em{\scshape i\kern-0.25em b}\kern-0.8em\TeX}}}

\copyrightyear{2024}
\acmYear{2024}
\setcopyright{acmlicensed}
\acmConference[DAC '24]{61st ACM/IEEE Design Automation Conference}{June 23--27, 2024}{San Francisco, CA, USA}
\acmBooktitle{61st ACM/IEEE Design Automation Conference (DAC '24), June 23--27, 2024, San Francisco, CA, USA}
\acmDOI{10.1145/3649329.3658266}
\acmISBN{979-8-4007-0601-1/24/06}

\acmPrice{}

\begin{document}

\title{UpDLRM: Accelerating Personalized Recommendation using Real-World PIM Architecture}



\author{Sitian Chen$^{1}$, Haobin Tan$^2$, Amelie Chi Zhou$^{1^*}$, Yusen Li$^3$, Pavan Balaji$^4$}
\affiliation{\institution{$^1$Hong Kong Baptist University\quad $^2$Shenzhen University \quad $^3$Nankai University\quad $^4$Meta}\country{}}\thanks{*Corresponding author: Amelie Chi Zhou}


\begin{abstract}
Deep Learning Recommendation Models (DLRMs) have gained popularity in recommendation systems due to their effectiveness in handling large-scale recommendation tasks. The embedding layers of DLRMs have become the performance bottleneck due to their intensive needs on memory capacity and memory bandwidth. 
In this paper, we propose \emph{UpDLRM}, which utilizes real-world processing-in-memory (PIM) hardware, UPMEM DPU, to boost the memory bandwidth and reduce recommendation latency. 
The parallel nature of the DPU memory can provide high aggregated bandwidth for the large number of irregular memory accesses in embedding lookups, thus offering great potential to reduce the inference latency. To fully utilize the DPU memory bandwidth, we further studied the embedding table partitioning problem to achieve good workload-balance and efficient data caching.
Evaluations using real-world datasets show that, UpDLRM achieves much lower inference time for DLRM compared to both CPU-only and CPU-GPU hybrid counterparts.
\end{abstract}






\maketitle

\vspace{-1ex}
\section{Introduction}

Deep Learning Recommendation Models (DLRMs) have become widely utilized for click-through rate (CTR) prediction and news rankings~\cite{naumov2019deep,covington2016deep}. According to Meta, DLRM models account for more than 60\% of their AI inference cycles in production, making them a primary target for optimization~\cite{gupta2020architectural}.
However, performance optimization for DLRMs is challenging, mainly due to their intensive needs on memory capacity and bandwidth~\cite{10137249}: 
\textbf{1)} The number of parameters in DLRMs is much larger than traditional deep learning models due to the \emph{embedding layers} which map high-dimensional categorical features (e.g., user IDs) to lower-dimensional dense vectors. The categorical features are encoded using embedding tables which can be very large in real-world scenarios (e.g., 10s of GBs). 
\textbf{2)} During inference, DLRMs perform multiple \emph{embedding lookups} and \emph{reductions}, requiring frequent and irregular memory accesses. These contribute significantly to the memory footprint and introduce memory bottlenecks, leading to increased latency.

Due to the limited memory capacity of GPUs, existing studies mainly adopt CPU-GPU hybrid architecture for DLRM inference, where the embedding tables (EMTs) are stored in CPU memories and GPUs are employed for highly parallel computations~\cite{naumov2019deep,ye2023grace}. Some recent work proposed to load popular data into limited GPU memory to further reduced data access latency~\cite{adnan2021accelerating}.
However, the hybrid architecture introduces unavoidable communication between CPUs and GPUs, and may cause GPU stalls due to the memory bottleneck of CPUs.  
Since DLRM inference is less compute-intensive compared to training, recent studies~\cite{10.1145/3579371.3589112} found it is possible to make real-time recommendations at scale based on only CPUs. However, memory is still a major performance bottleneck and workload-aware pre-processing (e.g., prefetching, caching) are needed to reduce inference latency~\cite{10.1145/3579371.3589112}.
In this paper, we proposed to boost the memory bandwidth of CPUs using real-world processing-in-memory (PIM) hardware, UPMEM DPU~\cite{upmemstat}.


UPMEM DPU is equipped with dedicated processing units connected closely to memory cells (details in Section~\ref{sec:pim}). We propose to store the large EMTs using DPU memory and perform costly memory lookups and reductions using DPUs. This design offers two compelling benefits.
\emph{First}, by offloading the memory-bound embedding operations to DPUs, we can reduce the resource contention on CPU memory bandwidth and hence accelerate the inference time. \emph{Second}, the parallel nature of DPU memory banks allows for efficient processing of multiple embedding lookups and reductions simultaneously, further accelerating the inference time. 

There exist several design issues for efficiently utilizing DPUs. Since the capacity of each DPU memory bank (MRAM) is limited to 64MB, it is usually necessary to partition an EMT onto multiple DPUs. 
When partitioning EMTs, we have to jointly consider the architectural features of DPUs and the memory access patterns of embedding operations to efficiently utilize the DPU memory bandwidth. \emph{First}, UPMEM DIMM does not support inter-DPU communication directly. Data movement across DPUs has to go through CPU memory. Thus, when partitioning EMTs, we should avoid frequent inter-DPU data exchange. \emph{Second}, each DPU core supports multi-threads and pipeline execution. We should fully utilize this hardware feature to improve the efficiency of DPU-supported embedding operations.
\emph{Third}, the popularities of items differ in real-world datasets~\cite{ye2023grace}. Thus, when items in EMTs are partitioned onto different DPUs, we have to consider the aggregated access frequency of each partition to achieve workload balance. 


In this paper, we propose \emph{UpDLRM}, which utilizes UPMEM PIM architecture to accelerate the memory-intensive embedding layers of DLRM, in order to achieve low inference time. Considering the above mentioned design issues, we study the EMT partitioning problem at three different levels. First, we consider the EMT partitioning only consider the hardware features, in order to fully utilize the memory bandwidth and maximize the throughput for embedding lookup operations (Section~\ref{sec:etp}). Second, we take the imbalanced data access frequencies into account, and proposed non-uniform EMT partitioning to achieve workload balance among DPUs. This ensures efficient utilization of DPU resources and minimizes embedding layer processing time (Section~\ref{sec:nup}). 
Third, we further consider the co-occurrence feature of multiple items in real recommendation systems, and adopt existing caching technique~\cite{ye2023grace} to further reduce memory traffic. To deal with the workload imbalance problem introduced by data caching, we proposed a cache-aware partitioning method that jointly balance the memory accesses on cache storage and regular EMT storage (Section~\ref{sec:cnup}).
Evaluations on six real-world datasets and four DLRM implementations using different hardware architectures have demonstrated the effectiveness of UpDLRM on optimizing the inference time. Overall, it achieved up to 4.6x speedup on inference time compared to the other comparisons.




\vspace{-1ex}
\section{Background and Motivation}

\subsection{DLRM Model}

\begin{figure}
    \centering
    \includegraphics[width=.72\linewidth]{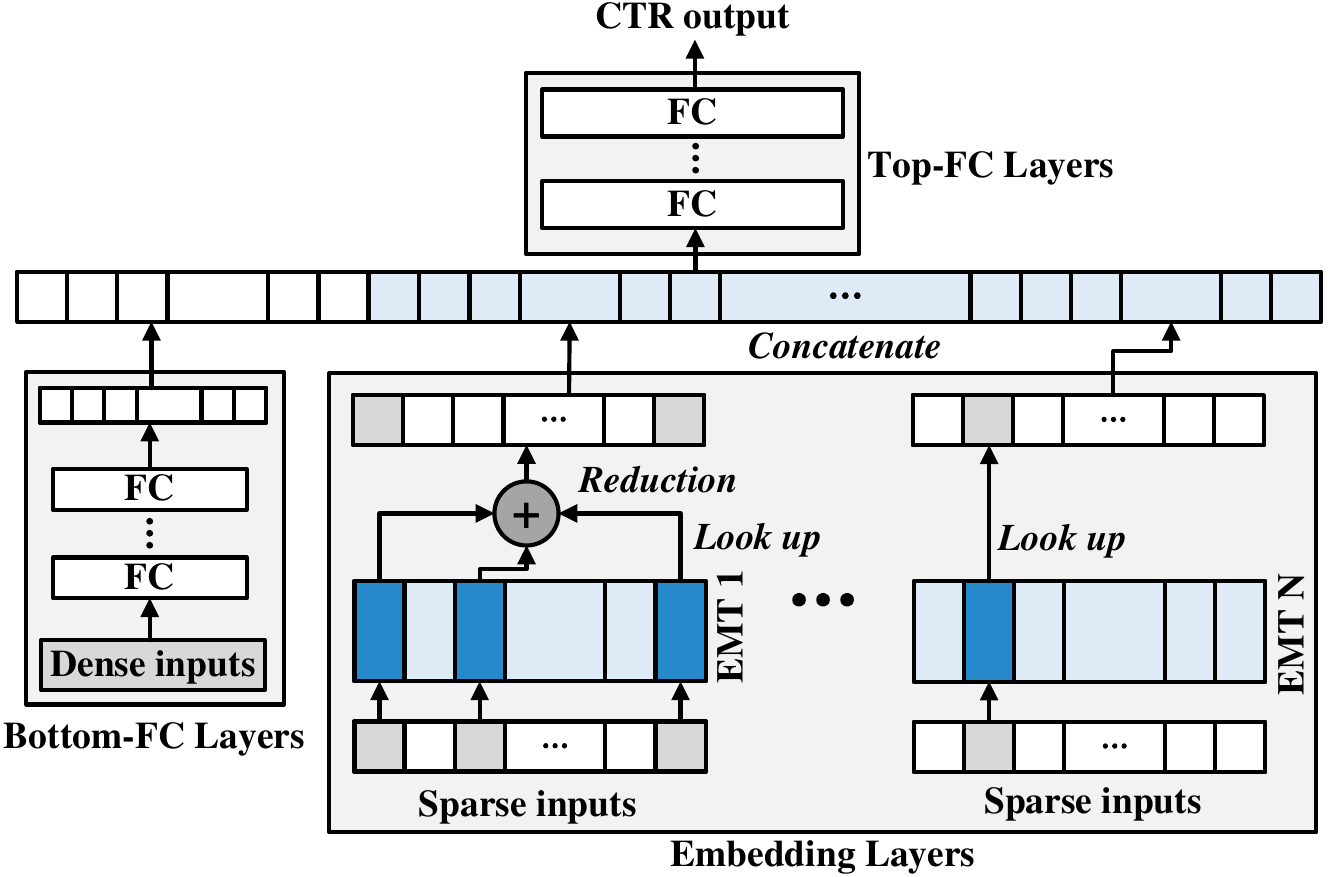}\vspace{-3ex}
    \caption{Architecture of DLRMs}
    \label{fig:dlrm}\vspace{-2ex}
    \vspace{-2ex}
\end{figure}

Figure~\ref{fig:dlrm} shows the structure of DLRMs. 
The bottom-FC layers process ``dense'' inputs that are usually continuous attributes (e.g., age, price, rating) to generate dense features. The embedding layers handle ``sparse'' inputs that are usually categorical features (e.g., user ID, gender, location) encoded using one-hot or multi-hot to generate embeddings. 
Each embedding layer contains a large embedding table (EMT) responsible for storing feature vectors of a specific category.
After obtaining the dense and sparse features, they are concatenated and fed into the top FC layer to predict the CTR.
The embedding layer is the main component that causes high memory capacity and bandwidth requirements of DLRMs. 

\emph{Memory Capacity Requirement:} The EMTs are essentially lookup tables, where each row in the table represents the embedding vector associated with a specific value of the category. Hence the size of a EMT is determined by the number of distinct values in a categorical feature. For example, consider the category being user ID. The number of users in a commercial recommendation system can easily reach billions (e.g., over 3 billions in Meta~\cite{naumov2020deep}). Assume the number of users is one billion and the embedding dimension is 20. The size of the table is $10^9*20*4$bytes, namely 80GB. Since there are usually many categories considered in real-world recommendation systems, the memory consumption of EMTs is substantial. 

\emph{Memory Bandwidth Requirement:} During DLRM inference, the embedding layer takes the input categorical values (indices) and performs \emph{a lookup operation} in the EMT to retrieve corresponding embedding vectors. This operation requires fetching embedding vectors from different memory locations based on the input indices, which can result in frequent memory accesses and require high bandwidth to sustain the data transfer rates.
In real-world scenarios, the popularity of different categorical values can vary significantly~\cite{ye2023grace}. This can lead to imbalanced accesses to different memory locations in the EMTs. 



\vspace{-2ex}
\subsection{UPMEM PIM Architecture}\label{sec:pim}
\begin{figure}
    \centering
    \begin{minipage}{.18\textwidth}
        \includegraphics[width=1\linewidth]{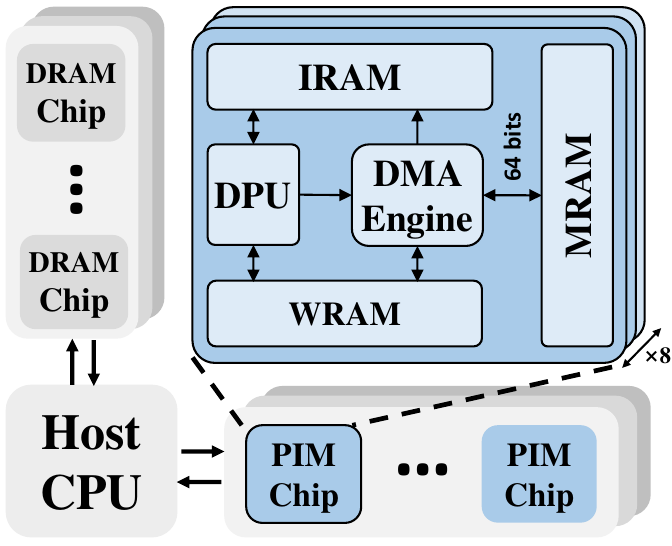}    \vspace{-3ex}
             \caption{UPMEM's PIM architecture}
         \label{fig:upmem}   
    \end{minipage}
    \hfil
    \begin{minipage}{.18\textwidth}
 \includegraphics[width=1\linewidth]{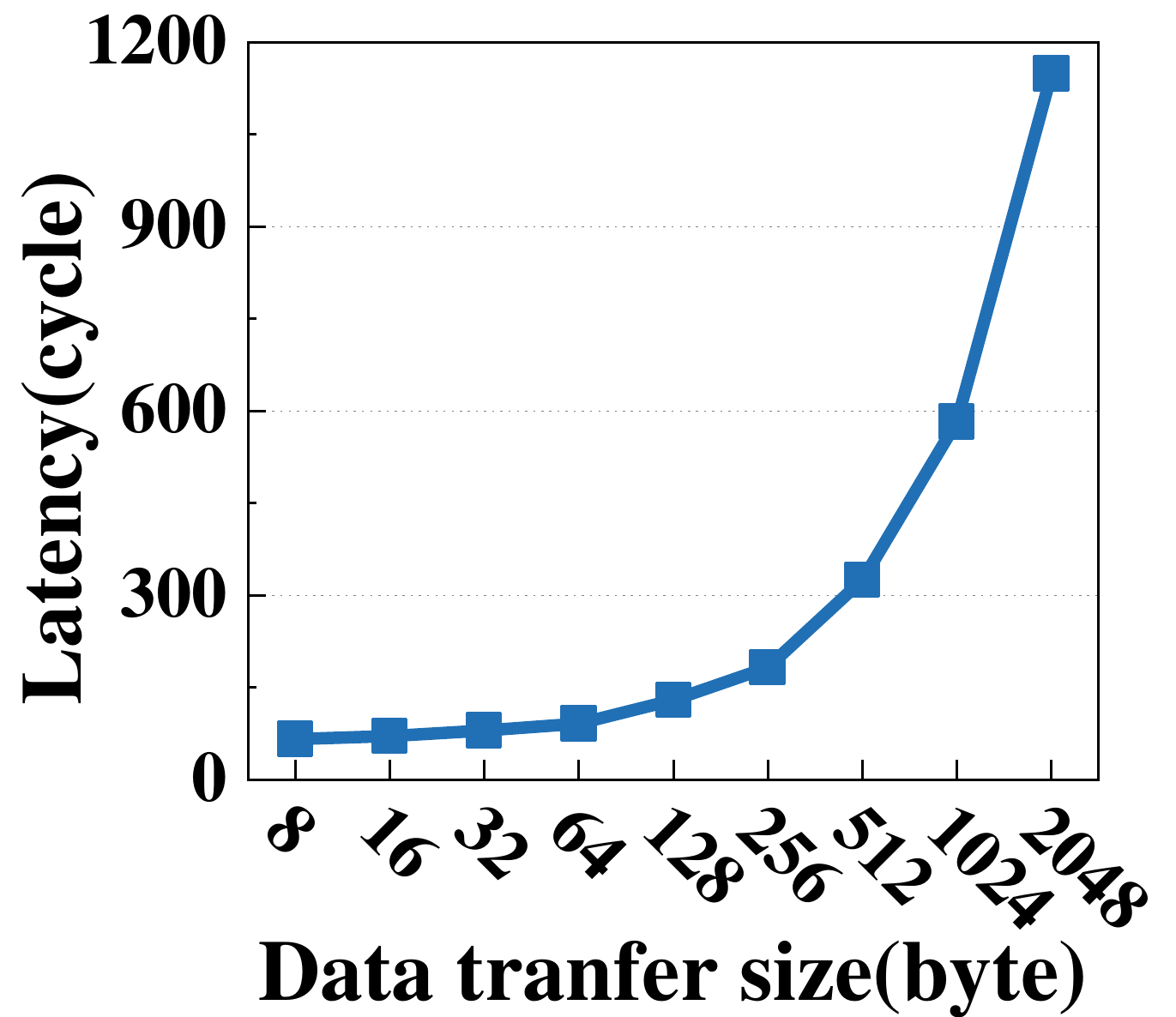}
    \vspace{-5ex}
    \caption{MRAM read latency}
    \label{fig:mram-latency}
    \end{minipage}
    \vspace{-3ex}
\end{figure}

The UPMEM PIM architecture is the first real-world PIM system commercialized in 2019~\cite{upmemstat,devaux2019true}. Figure~\ref{fig:upmem} plots the architecture of UPMEM's PIM system.
A UPMEM module consists of a standard DDR4 DIMM module that incorporates 8-16 PIM chips. Each chip comprises 8 processing cores known as DRAM Processing Units (DPUs). Each DPU has exclusive access to specific memory components, including a 64-MB DRAM bank MRAM, a 24-KB instruction memory IRAM, and a 64-KB scratchpad memory WRAM. The MRAM can be accessed by the host CPU for transferring input data (main memory to MRAM) and retrieving results (MRAM to main memory). These data transfers can occur concurrently if the buffers transferred to and from all MRAM banks are of the same size. Otherwise, the transfers happen sequentially. It's important to note that inter-DPU communication relies on the host CPU.

Different from RRAM-based PIM architectures~\cite{10244420,10137249}, which perform arithmetic operations using NOR gates within the RRAM arrays, UPMEM's PIM is the first ``real'' process in memory hardware that contains dedicated processing units, namely DPUs. Each DPU is a multi-threaded 32-bit RISC core with its own specific ISA~\cite{upmemusermanual} to facilitate data transfers between the MRAM and the WRAM. The maximum MRAM-WRAM bandwidth per DPU can achieve approximately 800 MB/s~\cite{upmemstat}. All DPUs in the UPMEM module function together as a parallel co-processor to the host CPU.

\vspace{-2ex}

\subsection{Motivation}

Embedding layers are the primary performance bottlenecks of DLRM inference due to their high requirements on memory capacity and memory bandwidth. The PIM architecture can be utilized to alleviate the memory bandwidth bottleneck and process embedding data near memory. 
Additionally, the UPMEM technical paper~\cite{upmemstat} indicates that the total cost of ownership (TCO) gain for the PIM platform is estimated to be around 10x, with a potential reduction of 60\% in energy consumption. Thus, studying PIM-accelerated DLRM inference could potentially lead to cost- and energy-efficient solution to industry.
In this paper, we propose to store the large embedding tables in the MRAM of DPUs and perform lookup operations using near memory DPU cores. 
Since each DPU MRAM has a limited capacity of 64MB, an EMT that is larger than 64MB has to be partitioned first to fit in the DPU memory. The partitioning has to consider several design issues as discussed next.

\vspace{-1ex}
\section{Design Details}

\begin{figure*}
    \centering
    \includegraphics[width=.7\linewidth]{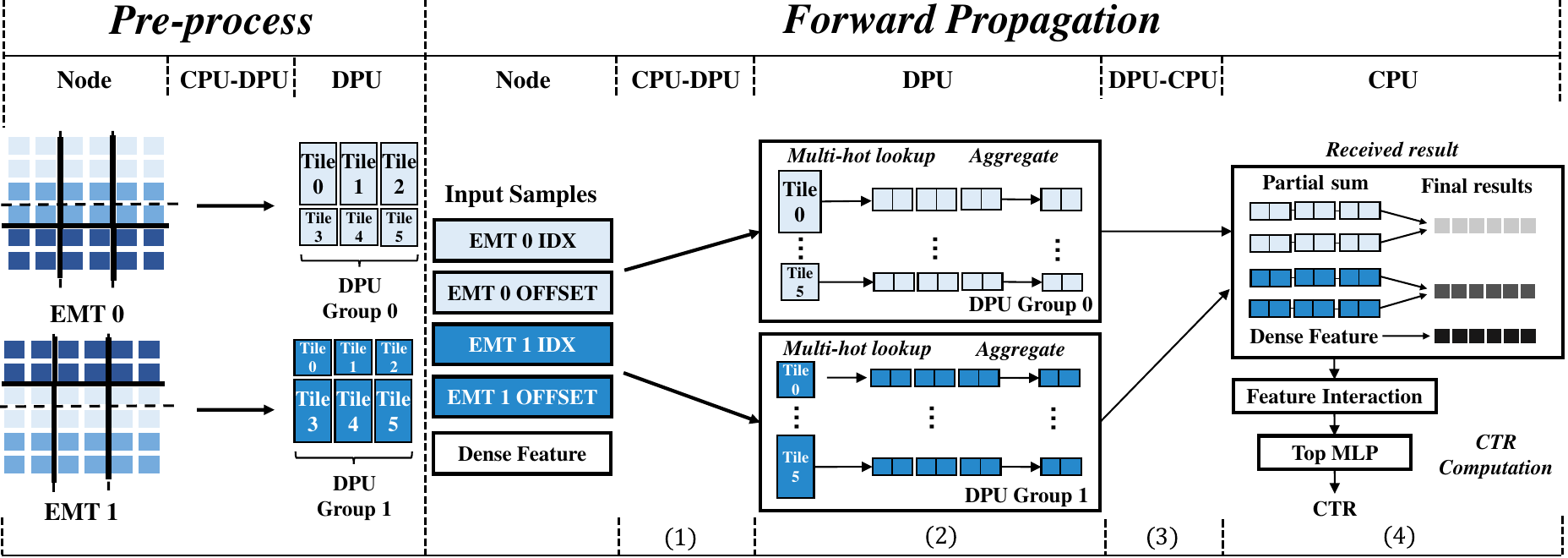}\vspace{-2ex}
    \caption{Pre-processing and forward propagation stages of our DPU-based DLRM implementation}
    \label{fig:dlrm-flow}
\end{figure*}

\begin{figure}
    \centering
    \vspace{-3ex}
    \begin{minipage}{.22\textwidth}
        \includegraphics[width=\linewidth]{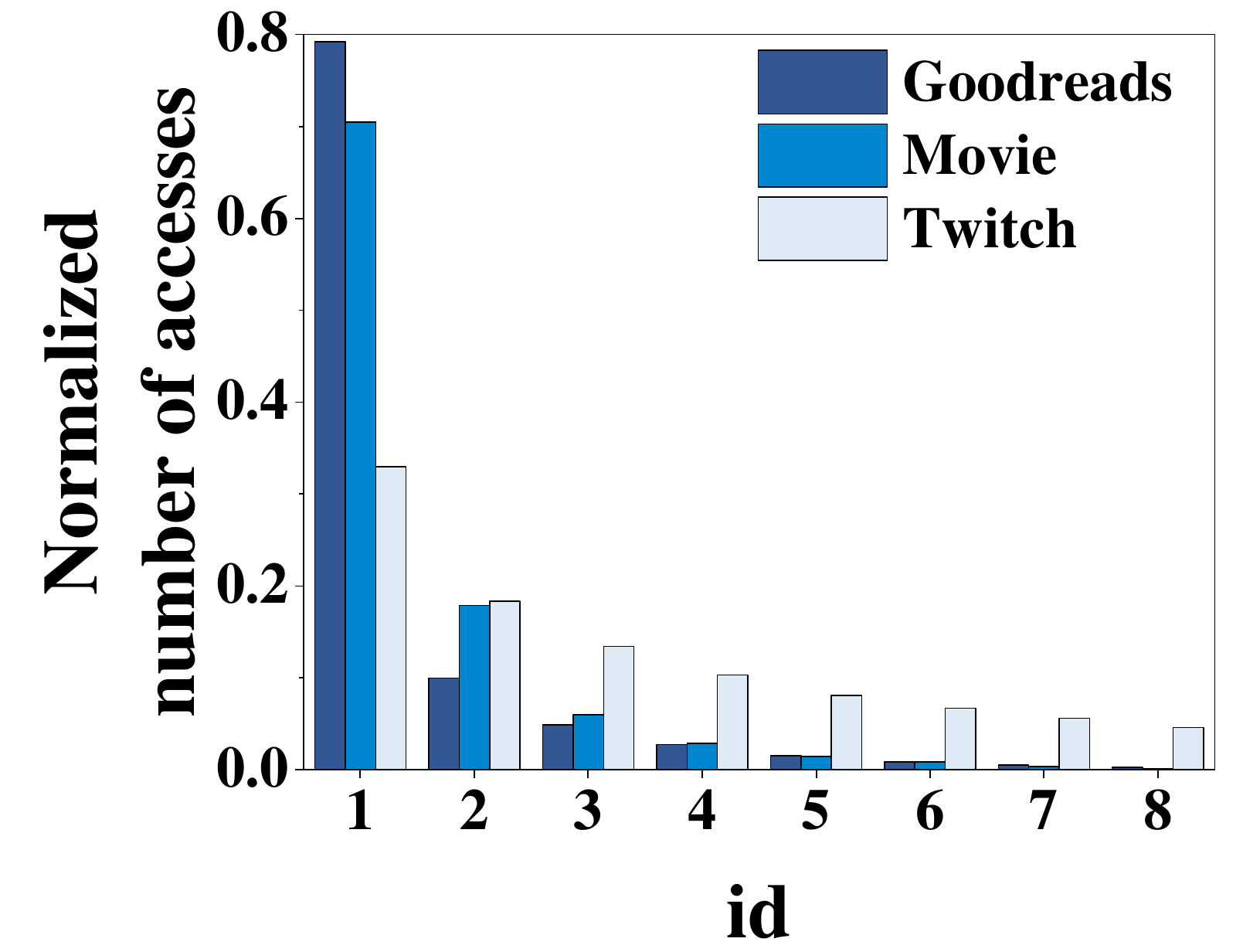}
    \vspace{-6ex}
    \caption{Proportion of partitions being accessed}
    \label{fig:partition}
    \end{minipage}
    \hfill
    \vspace{-3ex}
    \begin{minipage}{.22\textwidth}
    \includegraphics[width=\linewidth]{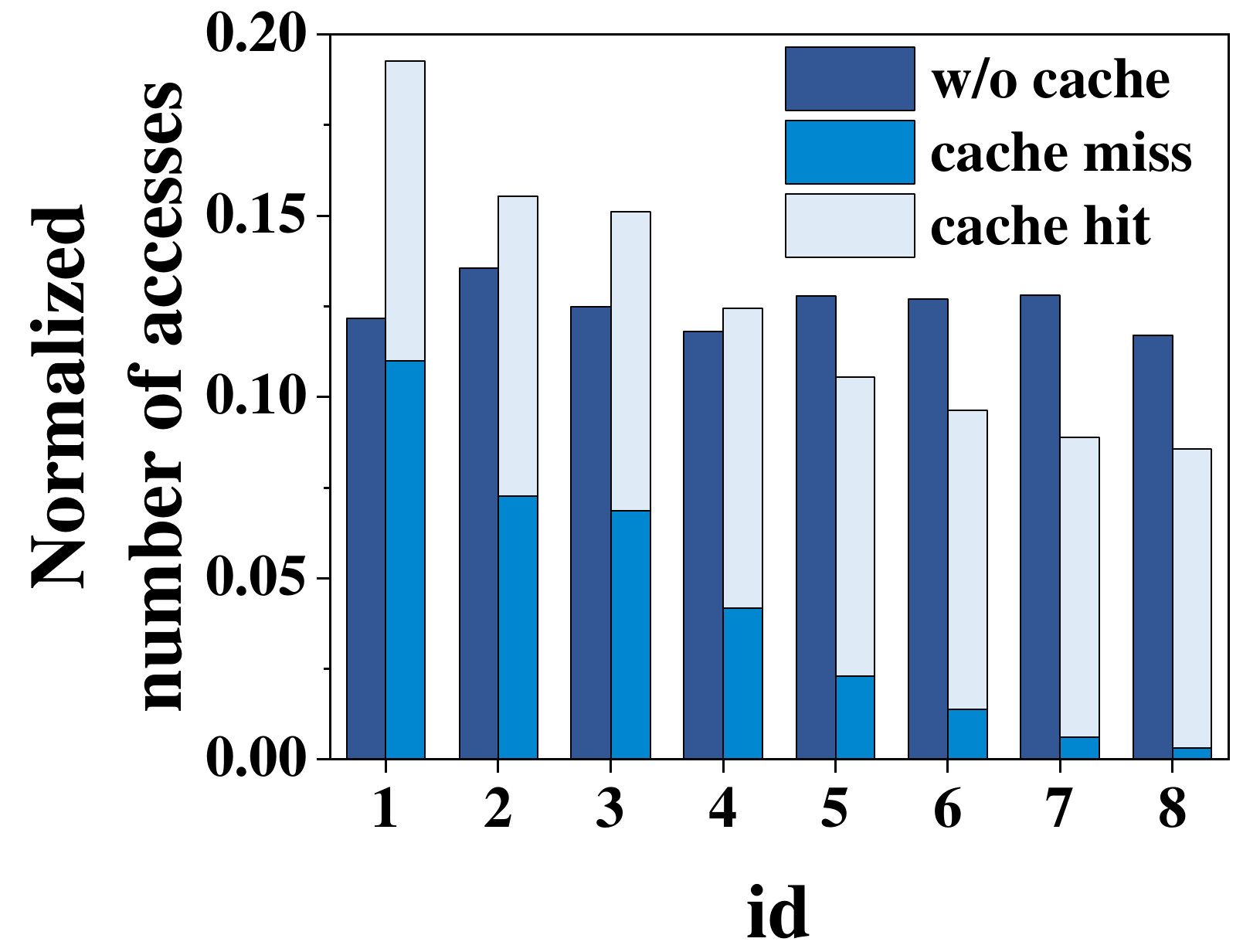}
        \vspace{-6ex}
    \caption{Access pattern w/ and w/o cache using Movie}
    \label{fig:powerlaw}
    \end{minipage}
\end{figure}

\subsection{Embedding Table Partitioning}\label{sec:etp}

We partition each embedding table into tiles smaller than 64MB to fit in the DPU memories. 
To simplify the problem, let's first consider uniform partitioning, namely all tiles have the same number of rows (denoted as $N_r$) and columns (denoted as $N_c$).
Assume each feature value is represented using 32-bit data, we have $N_r*N_c*$4B$\le 64$MB.

As shown in Figure~\ref{fig:dlrm-flow}, during the forward propagation, given a batch of embedding table indices and offsets (multi-hot vectors), the corresponding feature data are fetched from DPUs and transferred back to CPU.
The feature data fetching in stage 2 contains two steps. First, the multi-hot lookup operation searches in DPU MRAMs for corresponding feature data specified by IDX and OFFSET. 
Second, multiple vectors are returned and aggregated within DPU to utilize its computation power. 
Due to EMT partitioning, the aggregated embedding vectors are only partial results and need to be transferred back to the CPU to generate final feature vectors. We study the time spent on embedding layers as the sum of CPU-DPU communication time (stage 1), embedding lookup time (stage 2) and DPU-CPU communication time (stage 3).


Each embedding lookup operation reads $N_c$*4 bytes of data from MRAM at the same time. We study the performance of each memory access on MRAM as shown in Figure~\ref{fig:mram-latency}. Note that each MRAM read has to be 8 bytes aligned and can be 2,048 bytes maximum. Interestingly, we found that when the size of data increases from 8B to 32B, the latency increases rather slowly. After 32B, the latency increases more dramatically. The partitioning approach should carefully consider this feature when deciding the tile size. That is, we prefer to have $N_c$*4B $\le$ 32B, namely $N_c\le8$. 

Multiple DPUs perform lookup operations concurrently. Assuming balanced distribution of memory accesses in the dataset, uniform EMT partitioning can result in the same \emph{embedding lookup time} on each DPU. Approximately, we can estimate the lookup time on each DPU as 
$T_{lkp}=\frac{N_r}{R}*$\textit{batch\_size}$*$\textit{Avg\_Red}$*{t_a}$, where $t_a$ represents the memory access time indicated in Figure~\ref{fig:mram-latency} and \textit{Avg\_Red} represents the average reduction in the number of active features (ones) in the multi-hot encoding.
Similarly, the \emph{communication time} for transferring indices from CPU to DPU memory can be estimated as $T_{c-comm}=\frac{N_r}{R}*$\textit{batch\_size}$*$\textit{Avg\_Red}$*{t_c}$, where $t_c$ represents the time to transfer one value from CPU to DPU. The time for transferring partial results from DPU to CPU memory can be estimated as $T_{d-comm}=N_c*$\textit{batch\_size}$*t_d$, where $t_d$ represents the time to transfer one value from DPU to CPU. Thus, to minimize the time on embedding layers, we aim to find a good tradeoff between $N_r$ and $N_c$ to achieve the following goal: 
\begin{equation}
    \min T_{c-comm}+T_{lkp}+T_{d-comm}
\end{equation}
s.t., \vspace{-4ex}
\begin{equation}
    N_r*N_c =\frac{R*C}{N_{dpu}}\le 1.6*10^7 
\end{equation}\vspace{-2ex}
\begin{equation}
    N_c=2k, \quad 1\le k\le 4
    \end{equation}
\noindent where $N_{dpu}$ represents the number of DPUs used to stored EMTs. $R$ and $C$ represent the number of rows and columns of the EMT, respectively. Both $N_{dpu}$ and $R*C$ are fixed inputs to the problem.
The constraints greatly reduce the solution space of the problem. Thus we can simply search for the best $N_r$ and $N_c$ exhaustively.



\subsection{Non-Uniform EMT Partitioning}\label{sec:nup}
In the above discussion, we assume balanced distribution of memory accesses in a dataset, which is usually not true in real-world applications. 
We study the data access patterns of EMTs using three real-world datasets Goodreads~\cite{wan2018item,wan2019fine}, Movie~\cite{ni2019justifying} and Twitch~\cite{rappaz2021recommendation}. 
Figure~\ref{fig:partition} shows the total number of accesses per row block when dividing the rows into 8 blocks. Clearly, the three datasets all show imbalanced access distributions. The most popular block has 340 times higher number of accesses compared to the least popular block.
This will lead to workload imbalance problem to DPUs and make the uniform partitioning no longer optimal.

To accommodate this, we propose a non-uniform EMT partitioning method that considers the imbalanced data access pattern in real datasets. 
Specifically, consider using the same number of DPUs as the uniform partitioning, our non-uniform partitioning cuts an EMT into tiles of varying sizes and stores each tile in a separate DPU. DPUs used to store the same EMT collectively form a group as shown in Figure~\ref{fig:dlrm-flow}.
The goal is to achieve optimal workload balance among DPUs. We adopt a greedy method to achieve this goal. Specifically, by profiling the historical user-item access trace, we can obtain the access frequencies of all items. Consider each DPU as a bin, the problem of assigning each row to DPUs is a classical Bin-Packing problem with a fixed number of bins. We first order items according to their frequencies and then iteratively assign each row to the bin with the lowest total frequency unless the bin is full (64MB capacity). Note that we adopt the same $N_c$ as optimized in uniform partitioning.
The complexity of the algorithm is hence $O(R)$. One could batch items when doing the assignment to reduce algorithm complexity.
Consider the Movie dataset which has a skewed item access frequency as shown in Figure~\ref{fig:partition}.
When partitioning the EMT into 8 row blocks with our non-uniform algorithm, it can be observed from Figure~\ref{fig:powerlaw} that the number of accesses per partition is much more balanced (i.e., w/o cache).

\floatname{algorithm}{Algorithm} 
\renewcommand{\algorithmicrequire}{\textbf{Input:}}  
\renewcommand{\algorithmicensure}{\textbf{Output:}}  
\begin{algorithm}[t]\scriptsize
	\caption{Cache-Aware Non-Uniform Partitioning}\label{alg:pnup}
	\begin{algorithmic}[1]
		\Require 
		{$N_{dpu}$: The number of DPUs/partitions;\newline 
            $obj\_freq$: The item access frequency trace;\newline	
		$cache\_res$: A list of cached partial sums generated by GRACE~\cite{ye2023grace}.}
		\Ensure {$assigned\_part$: Assigned partition of each item.}
  
        \State Sort $obj\_freq$ in descending order;
        \State $p\_id = 0$;
        \State $part\_count$ counts the current size of each partition; 
        \For{$list$ in $cache\_res$ } \Comment{Cache hit}
            \State $benefit = list[-1]$;
            \Comment{Reduced memory access estimated by GRACE}
            \State $p\_id$ $\leftarrow$ The partition with the lowest $part\_count$ value and enough cache capacity;
            \For{$index$ in $list$}
                \State $assigned\_part[index] = p\_id$;
                \State $part\_count[p\_id] += obj\_freq[index]$;
            \EndFor
            \State $part\_count[p\_id] -= benefit$;
        \EndFor
        \For{$index$ in $obj\_freq$} \Comment{Cache miss}
            \If{$index \notin cache\_res$}
               \State $p\_id$ $\leftarrow$ The partition with the lowest $part\_count$ value and enough EMT capacity;
               \State $assigned\_part[index] = p\_id$;
               \State $part\_count[p\_id] += obj\_freq[index]$;
            \EndIf
        \EndFor
        \State \textbf{return} $assigned\_part$;
	\end{algorithmic}
\end{algorithm}


\vspace{-1ex}
\subsection{Cache-Aware Non-Uniform Partitioning}\label{sec:cnup}

\begin{figure}[t]
    \vspace{-1ex}
    \centering
    \includegraphics[width=.8\linewidth]{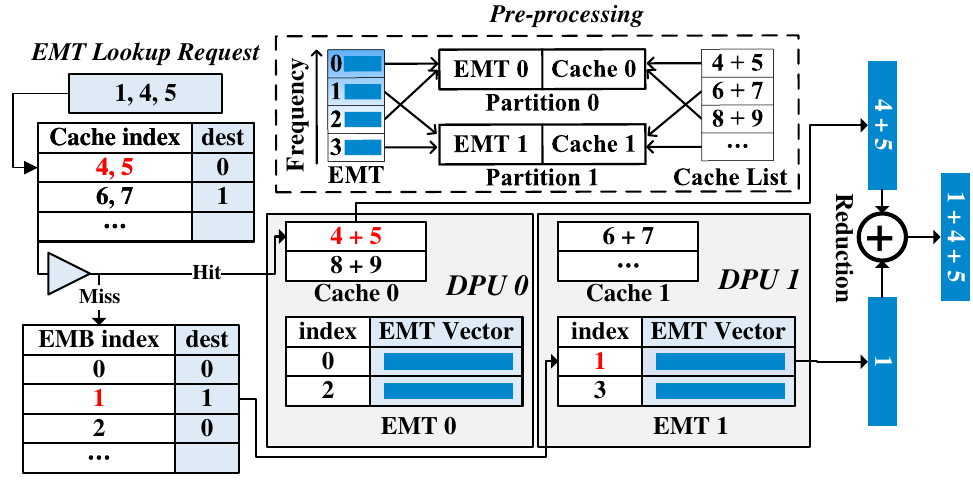}\vspace{-3ex}
    \caption{Example of cache-aware partitioning}
    \label{fig:pa:example}\vspace{-4ex}
\end{figure}

Existing studies~\cite{ye2023grace,kal2021space} leverage the power-law distribution of item access frequencies to cache the partial sum of popular items to reduce memory traffic and accelerate DLRM inference.
For example, the recent GRACE~\cite{ye2023grace} algorithm takes the \emph{co-occurrences} of popular items into account, and adopts a graph-based algorithm to identify frequently accessed item combinations. These combinations' partial sums are cached to reduce memory traffic. 
The caching techniques have already demonstrated effectiveness on reducing DLRM inference latency. However, when applied to our DPU-based DLRM, they further exacerbate the workload imbalance problem among DPUs. 
As shown in Figure~\ref{fig:powerlaw}, when partitioning the Movie dataset using our non-uniform algorithm with GRACE-based caching, the total number of memory accesses is reduced by 40\% compared to without caching, while the access pattern is much imbalanced across DPUs.

To address this issue, we propose cache-aware non-uniform partitioning as shown in Algorithm~\ref{alg:pnup}. 
The function takes three inputs, including $N_{dpu}$, $obj\_freq$ which records the access frequency history of different items, and $cache\_res$ which is a cache list containing partial sums generated by existing caching technique~\cite{ye2023grace}. A cache list of \{$a,b,c$\} means that items $a$, $b$ and $c$ frequently co-exist in the same sample, so that partial sums $a$, $b$, $c$, $a+b$, $a+c$, $b+c$ and $a+b+c$ are cached to reduce the memory traffic. We divide each MRAM capacity into two parts: one storing embedding vectors and the other storing the cached partial sums. 
Initially, we sort $obj\_freq$ in descending order. That is, items with higher frequencies are ranked in front. To decide the partition of an item, we first look for it in the cache list (Line 4). If it exists in the list, we store the item in a partition with the minimum total frequency and has enough capacity for cache storage (Line 5-10). If the item does not exist in cache list, we store it in the EMT part of MRAM storage (Line 12-15). With this greedy algorithm, we aim to balance the combined number of memory accesses (EMT+cache) per partition.



Figure~\ref{fig:pa:example} illustrates an example of our cache-aware partitioning algorithm. Consider a sample request to look up indices 1, 4 and 5. The system checks the cache index to determine if the indices are stored in the cache space. For example, in our scenario, indices 4 and 5 are stored in DPU0's cache space, allowing us to read the MRAM once to obtain the partial sum 4 + 5. This helps to reduce the number of memory accesses from two to one. Index 1 is not stored in the cache space, so we retrieve its embedding vector from the EMT space. Ultimately, each DPU returns a partial sum to the CPU (e.g., 4+5 and 1). We combine and aggregate these partial sums using CPU to obtain final results.


In the above design, one parameter is the cache capacity. The cache space and EMT space share the 64MB MRAM size. The larger share for cache space, the higher reduction can be obtained on embedding lookup time. For example, using GoodReads dataset and varying the cache capacity to 40\%, 70\% and 100\% of the required storage size of cache list, our cache-aware non-uniform partitioning is able to reduce the embedding lookup time by 17\%, 22\% and 26\%, respectively, compared to that of without caching. The downside of storing large cache list is that it requires larger memory capacity. By default, we set the cache capacity to 100\% of the required size.

\vspace{-1ex}
\section{Evaluation}

\begin{table}[]
\vspace{-3ex}
\caption{Workload Configurations}\label{tab:workload}
\vspace{-3ex}
\resizebox{\linewidth}{!}{%
\begin{tabular}{c|ccc}
\hline
\textbf{\textbf{\textbf{Category}}} &
  \textbf{\textbf{\textbf{\textbf{Workload}}}} &
  \textbf{\textbf{\textbf{\textbf{Avg.Reduction}}}} &
  \textbf{\textbf{\textbf{\textbf{$\#$Items}}}} \\ \hline
\multirow{2}{*}{Low Hot}  & AmazonClothes(clo)~\cite{ni2019justifying}       & 52.91  & 2,685,059 \\
                          & AmazonHome(home)~\cite{ni2019justifying}                                 & 67.56  & 1,301,225 \\ \hline
\multirow{2}{*}{Medium Hot}                & MetaFBGEMM1(meta1)~\cite{metadatasets}           & 107.2  & 5,783,210 \\
                          & MetaFBGEMM2(meta2)~\cite{metadatasets}           & 188.6  & 5,999,981 \\ \hline
\multirow{2}{*}{High Hot} & GoodReads(read)~\cite{wan2018item,wan2019fine}   & 245.8  & 2,360,650 \\
                          & GoodReads2(read2)~\cite{wan2018item,wan2019fine} & 374.08 & 2,360,650 \\ \hline
\end{tabular}%
}
\end{table}

\subsection{Experimental Setup}

\textbf{Workloads.} We adopt Meta's deep learning recommendation model (DLRM)~\cite{naumov2019deep} with six real-world datasets\footnote{\url{https://cseweb.ucsd.edu/~jmcauley/datasets/amazon\_v2}\\\url{https://mengtingwan.github.io/data/goodreads.html}}, as shown in Table~\ref{tab:workload}. The datasets can be categorized into three groups, namely low hot, medium hot and high hot, according to the average reduction frequency of the dataset. 
In our experiments, we duplicate each dataset to form eight EMTs and each embedding vector has 32 dimensions. To measure inference performance, we conduct a sampling of 12,800 inferences in each set of experiments. The batch size is set to 64. 

\textbf{Comparisons.} We compare UpDLRM with three other open-source DLRM implementations, as detailed in Table~\ref{tab:hardware-env}. 
DLRM-CPU is the CPU-only implementation that adopts CPU for both EMT storage and parallel computations. DLRM-Hybrid and FAE~\cite{adnan2021accelerating} are two implementations based on the CPU-GPU hybrid architecture. In the hybrid architecture, CPU is used to store EMTs and perform embedding lookups, while the GPU manages CTR computation. Embedding lookup results are communicated to the GPU via PCI-e. The FAE approach differs from DLRM-Hybrid in that it involves placing a subset of highly accessed item embedding vectors in a cache space, such as the GPU memory, to accelerate DLRM inference time.
Our UpDLRM implementation is illustrated in Figure~\ref{fig:dlrm-flow}, where cache-aware EMT partitioning is used in the pre-process stage. We adopt two UPMEM modules, totalling 256 DPUs. Each DPU employs 14 tasklets. 


\begin{table}[]
\caption{Specifics of evaluated hardware architectures}\label{tab:hardware-env}
\vspace{-2ex}
\resizebox{\linewidth}{!}{%
\begin{tabular}{c|c|c|c}
\hline
\textbf{\textbf{\textbf{Implementation}}} & \textbf{\textbf{\textbf{Architecture}}} & \textbf{\textbf{\textbf{CPU core}}} & \textbf{\textbf{\textbf{Memory}}} \\ \hline
DLRM-CPU~\cite{naumov2019deep}                & Intel Xeon(R) Silver 4110(2.10GHz) & 32 & 128GB        \\ \hline
DLRM-Hybrid~\cite{adnan2021accelerating}             & Intel Xeon(R) Silver 4110(2.10GHz) & 32 & 128GB        \\ \cline{1-1}
FAE~\cite{adnan2021accelerating}                    & NVIDIA GeForce GTX 1080 Ti         & -  & 11GB         \\ \hline
\multirow{2}{*}{UpDLRM} & Intel Xeon(R) Silver 4110(2.10GHz) & 32 & 128GB        \\
                        & UPMEM DPU(350MHz) $\times$ 256                 & -  & 16GB \\ \hline
\end{tabular}%
}
\end{table}

\begin{figure}
    \centering
    \includegraphics[width=.75\linewidth]{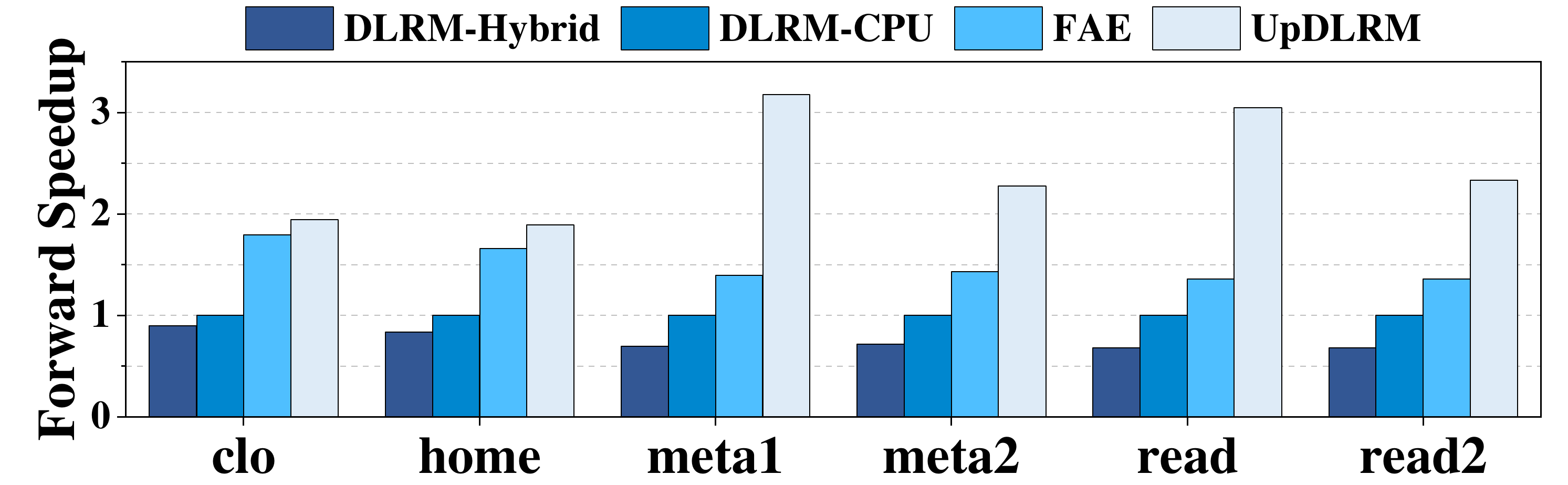}\vspace{-3ex}
    \caption{Inference performance speedup of compared approaches over that of DLRM-CPU}
    \label{fig:infer:res}
    \vspace{-3ex}
\end{figure}
\begin{figure}
    \centering
    \includegraphics[width=.85\linewidth]{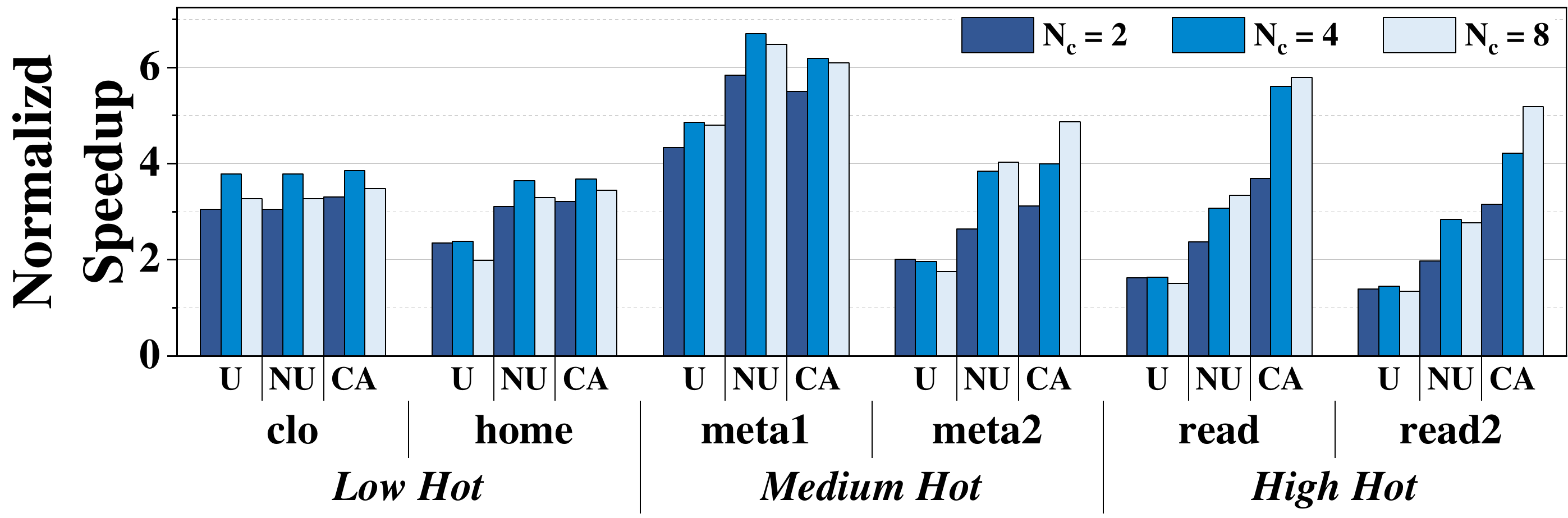}\vspace{-3ex}
    \caption{Performance speedup of embedding layers obtained by three partitioning methods over DLRM-CPU}
    \label{fig:emb:res}
    \vspace{-3ex}
\end{figure}

\vspace{-1ex}
\subsection{Effectiveness of UpDLRM}\label{sec:tmp}
Figure~\ref{fig:infer:res} illustrates the performance of the compared implementations on the six datasets, in terms of inference time speedup compared to that of DLRM-CPU. 
Overall, UpDLRM obtains the best speedup performance among all comparisons for all datasets. Specifically, UpDLRM accelerate the inference time by 1.9x-3.2x, 2.2x-4.6x and 1.1x-2.3x compared to DLRM-CPU, DLRM-Hybrid and FAE, respectively. Especially, it obtains higher speedup when the average reduction of the dataset is high. A larger average reduction represents a higher number of embedding lookup operations in a dataset. Thus, the results demonstrate that our DPU-based architecture is effective on improving the performance of embedding layers.
Unsurprisingly, FAE outperforms DLRM-CPU and DLRM-Hybrid, demonstrating the effectiveness of its caching technique.
DLRM-Hybrid performs the worst among all, mainly due to the fact that significant time is required by the CPU to execute embedding lookups, leading to GPUs waiting for the embedding results before proceeding with CTR computation.



To learn the improvements obtained by UpDLRM in details, we further perform comparative evaluation of the three proposed EMT partitioning methods, namely uniform (U), non-uniform (NU) and cache-aware (CA), using the six datasets. Figure~\ref{fig:emb:res} illustrates the performance speedup of the embedding layer obtained by the three methods compared to DLRM-CPU. The number of columns per partition (i.e., $N_c$) is fixed to 2, 4 or 8 for different partitioning methods.
We have the following observations: 1) Under the same $N_c$ value, CA partitioning obtains larger speedup compared to U and NU for High Hot datasets. This explains the good performance of UpDLRM under High Hot in Figure~\ref{fig:infer:res}.
2) The three partitioning methods perform almost the same for the ``clo'' dataset. This is because the item access pattern in this dataset is quite balanced, and the cache rate is low.
3) There's no universally good choice of $N_c$ for different datasets. For example, $N_c=4$ works the best for the first three datasets while $N_c=8$ is optimal for the latter three. Recall that a large $N_c$ value leads to higher DPU-CPU communication time but lower embedding lookup time and CPU-DPU communication time, as discussed in Section~\ref{sec:etp}. UpDLRM can automatically achieve good balance between the two for different datasets.

\vspace{-1ex}
\subsection{Latency breakdown}

\begin{figure}
    \centering
    \begin{minipage}{.22\textwidth}
        \includegraphics[width=.9\linewidth]{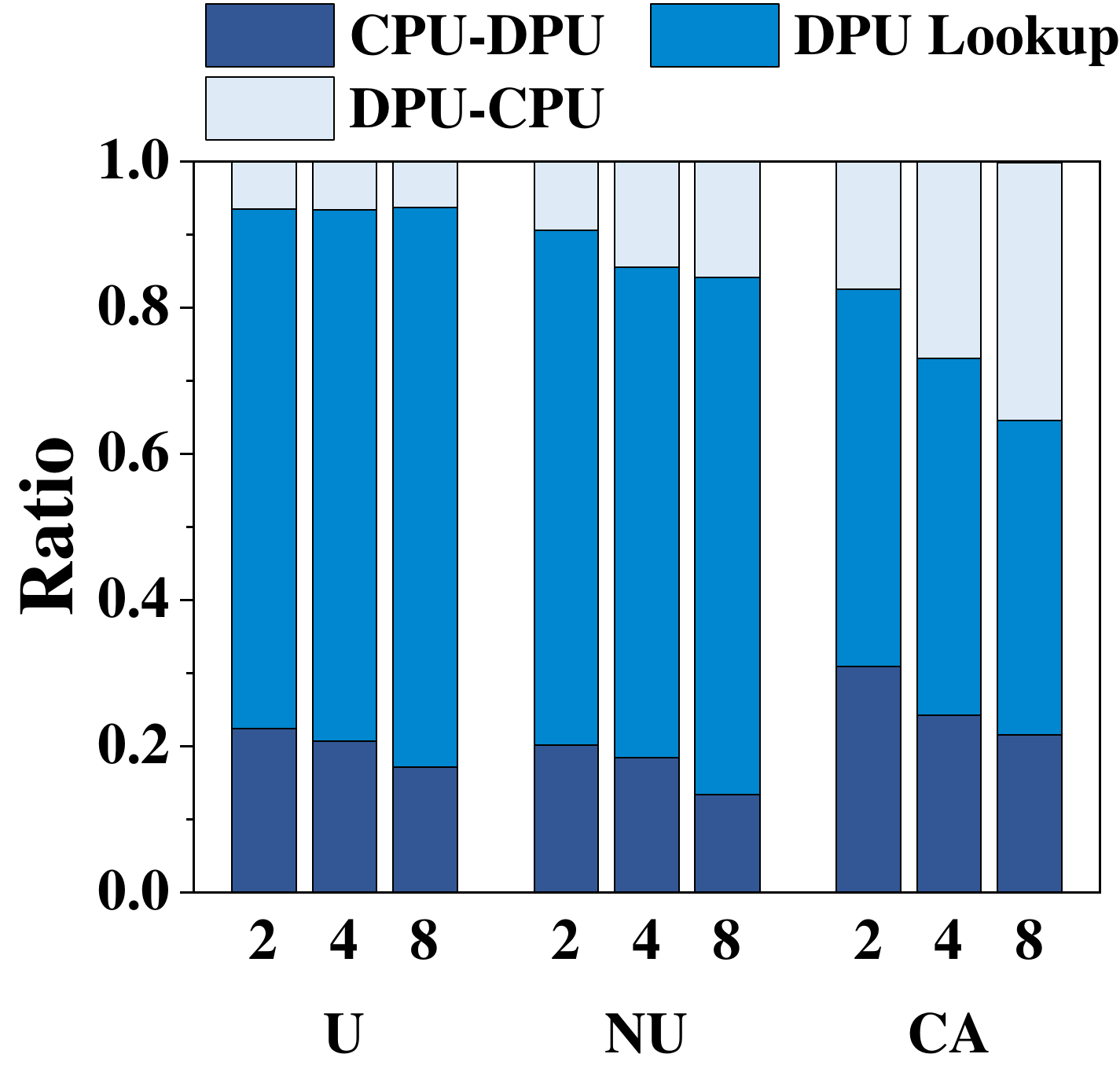}
     \vspace{-3ex}
    \caption{Latency breakdown of embedding layers with three partitioning methods using Goodreads. $N_c$ is fixed at 2, 4 or 8.}
    \label{fig:breakdown:res}
    \end{minipage}
    \hfill
    \begin{minipage}{.22\textwidth}
    \vspace{4ex}
    \includegraphics[width=.9\linewidth]{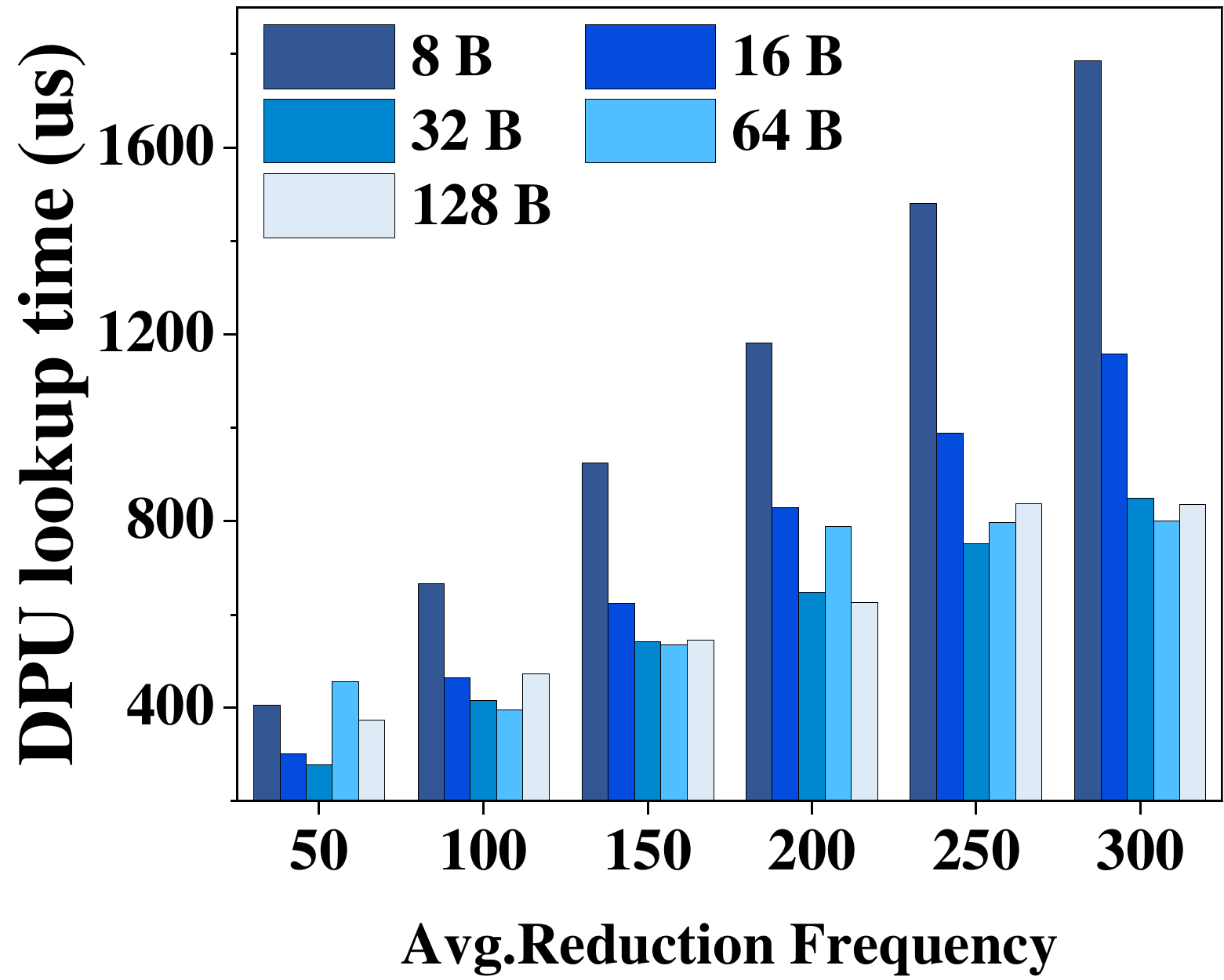}
        \vspace{-3ex}
    \caption{The DPU lookup time under varying average reduction frequencies and lookup data sizes (8B to 128B).}
    \label{fig:exp:sensitivity}
    \end{minipage}
    \vspace{-3ex}
\end{figure}

To study the performance of each component of Figure~\ref{fig:dlrm-flow}, we further breakdown the latency of UpDLRM into three parts, namely CPU-DPU communication time (stage 1), DPU lookup time (stage 2) and DPU-CPU communication time (stage 3). 
Figure~\ref{fig:breakdown:res} shows the breakdown results when using uniform (U), non-uniform (NU) and cache-aware (CA) partitioning in pre-process stage of UpDLRM. $N_c$ is fixed at 2, 4 and 8. We have the following observations.

First, comparing different partitioning methods, CA reduces the ratio of DPU lookup time in the embedding time from 71\%-77\% to 43\%-52\%, compared to the other partitioning methods. This means that the partial sum caching together with the cache-aware partitioning are effective on reducing the embedding lookup time in UpDLRM. With the ratio of embedding lookup time decreases to below 50\%, it means the bottleneck effect of the embedding operations has been greatly mitigated.

Second, comparing the same partitioning method under different $N_c$ values, the CPU-DPU communication time increases and DPU-CPU communication time decreases with the increase of $N_c$. For example, for CA, the CPU-DPU time ration reduces from 31\%-21\% and DPU-CPU time ratio increases from 17\%-35\% when $N_c$ increases from 2 to 8. This is consistent with our analysis of UpDLRM performance in Section~\ref{sec:etp}.




\vspace{-1ex}
\subsection{Sensitivity Study}
In Figure~\ref{fig:infer:res}, we have observed better performance of UpDLRM for datasets with higher average reduction.
To better study the potential of UpDLRM, we create a set of synthetic datasets with balanced item access patterns and varying average reduction frequencies ranging from 50 to 300. We vary $N_c$ from 2, 4, ..., to 32, thus resulting in varying data sizes of 8B, 16B, ..., to 128B. These sizes correspond to the amount of data loaded from MRAM per lookup. We set batch size to 64 samples.
Figure~\ref{fig:exp:sensitivity} shows the DPU lookup time results under different configurations.
We have the following observations.

First, under the same lookup data size, the DPU lookup time increases almost linearly with the increase of average reduction frequency. For example, when the lookup data size is 8B, the DPU lookup time increases from 406us to 1786us. When the data size gets larger, the increase of DPU lookup time with the increase of average reduction becomes rather stable. For example, when the lookup data size is 64B, the DPU lookup time increases from 456us to 787us when the average reduction increases from 50 to 200, and remains almost unchanged when the average reduction continues to increase to 300. This is because we employ 14 tasklets in one DPU, and the pipeline in high hot scenarios effectively masks the MRAM read latency, resulting in a comparable latency. The results demonstrate that, the embedding lookup performance of UpDLRM is quite scalable with the increase of average reduction frequency.

Second, given a fixed average reduction frequency, the DPU lookup time greatly reduces when the lookup data size increases from 8B to 32B. Recall our previous observation in Figure~\ref{fig:mram-latency}, the MRAM access latency remains almost unchanged when the access data size varies between 8B and 32B. With larger per lookup data size, less number of lookup operations is needed and thus lower lookup time can be obtained. When the access data size increases to larger than 32B, the per lookup latency also increases, thus hinders the reduction of DPU lookup time.
Thus, in previous experiments, we only set $N_c$ to 2, 4 and 8 to study the performance of UpDLRM.



\vspace{-2ex}
\section{Related work}
PIMPR~\cite{10137249} utilizes ReRAM crossbar, using NOR gates within the ReRAM arrays to maintain the Computing-In-Memory capability, to enhance DLRM acceleration. However, its hardware structure differs from UPMEM PIM, the first commercial processing-in-memory chip. RecNMP~\cite{ke2020recnmp} implements near-memory processing of the embedding vector around specialized DIMMs to expand the accelerator's memory space to tens of GBs. These studies are orthogonal to ours due to the different hardware architectures. 
EVStore~\cite{kurniawan2023evstore} implements a three-degree cache method to store popular items. It utilizes mixed precision datatype to cache a larger number of embedding vectors and leverages the similarity between embedding vectors to further enhance the cache hit rate. In addition, GRACE~\cite{ye2023grace} and SPACE~\cite{kal2021space} leverage hot items to reduce memory access, caching the hot embedding vector and partial sum in a cache space. However, they do not consider scenarios involving allocating the hot embedding vector in multiple cache spaces, such as caching in multiple DPUs, potentially leading to workload imbalance. UpDLRM introduces a cache allocation method across multiple cache spaces and addresses the workload imbalance issue arising after caching. Note that, although we adopt GRACE to generate cache list in this paper, UpDLRM does not rely on GRACE and can work with any other caching technique.

\vspace{-1ex}
\section{Conclusion}
This paper proposes \emph{UpDLRM}, which utilizes real-world processing-in-memory (PIM) hardware, UPMEM DPU, to boost the memory bandwidth and reduce recommendation latency. The parallel nature of the DPU memory can provide high aggregated bandwidth for the large number of irregular memory accesses in embedding lookups, thus offering great potential to reduce the inference latency of DLRMs. To fully utilize the DPU memory bandwidth, we further studied the embedding table partitioning problem to achieve good workload-balance and efficient data caching.
Evaluations using real-world datasets show that, UpDLRM achieves up to 4.6x speedup in terms of inference performance compared to both CPU-only and CPU-GPU hybrid counterparts. In the future, we plan to work on designing a DPU-GPU heterogeneous system to further optimize the inference time of DLRM systems.

\vspace{-1ex}
\section*{Acknowledgement}
This work is supported by National Natural Science Foundation of China (62172282, 62293510, 62293513, 62272252, 62272253), Guangdong Natural Science Foundation 2022A1515010122, Shenzhen Science and Technology Foundation RCYX20221008092908029, and a startup grant of HKBU.

\vspace{-1ex}
\bibliographystyle{ACM-Reference-Format}
\bibliography{main}

\end{document}